\documentclass[twocolumn,prl,showpacs,superscriptaddress]{revtex4}
\usepackage{graphicx}
\usepackage{natbib}
\usepackage{latexsym}
\usepackage{amsmath}
\begin{document}

\newcommand{\dum}{$\left.\right.$}

\title
{ Relaxation in a Completely Integrable Many-Body Quantum System: An
{\it Ab Initio} Study of the Dynamics of the Highly Excited States
of 1D Lattice Hard-Core Bosons }
\author{Marcos Rigol}
\affiliation{Permanent Address: Physics Department, University of
California, Davis, CA  95616, USA}
\author{Vanja Dunjko}
\affiliation{Permanent Address: Department of Physics \& Astronomy,
University of Southern California, Los Angeles, CA 90089, USA}
\affiliation{Institute for Theoretical Atomic and Molecular Physics, 
Cambridge, MA 02138, USA}
\author{Vladimir Yurovsky}
\affiliation{Permanent Address: School of Chemistry, Tel Aviv University,
Tel Aviv 69978, Israel}
\author{Maxim Olshanii}
\email{olshanii@physics.usc.edu}
\affiliation{Permanent Address: Department of Physics \& Astronomy,
University of Southern California, Los Angeles, CA 90089, USA}
\affiliation{Institute for Theoretical Atomic and Molecular Physics, Cambridge, MA 02138, USA}
\date{\today }

%
\begin{abstract}
In this Letter we pose the question of whether a many-body
quantum system with a full set of conserved quantities 
can relax to an equilibrium state, and, if it can, 
what the properties of such state are. 
We confirm the relaxation hypothesis through a thorough 
{\it ab initio} numerical investigation of the dynamics of 
hard-core bosons on a one-dimensional lattice. Further, a 
natural extension of the Gibbs ensemble to integrable systems 
results in a theory that is able to predict the mean values
of physical observables after relaxation.
Finally, we show that our generalized equilibrium carries 
more memory of the initial conditions than the usual 
thermodynamic one. This effect may have many experimental 
consequences, some of which having already been observed 
in the recent experiment on the non-equilibrium dynamics 
of one-dimensional hard-core bosons in a harmonic potential 
[T. Kinoshita, T. Wenger, D. S. Weiss, 
Nature (London) {\bf 440}, 900 (2006)].
\end{abstract}

\maketitle

{\it Introduction}.--
Integrable quantum gases traditionally belong to the domain of
mathematical physics, with little or no connection to experiments.
However, the experimental work on confined quantum-degenerate
gases has recently yielded faithful realizations of a number of
integrable one-dimensional many-body systems, thus making them
phenomenologically relevant. Examples include the gas of hard-core
bosons (Girardeau) \cite{Imp_Bosons,Lenard, Vaidya}, realized in
\cite{Weiss_1D}; its lattice version \cite{rigol04_1} 
studied in our Letter, realized in \cite{Paredes_Nature}; 
finite-strength $s$-wave-interacting spin-$0$ bosons 
(Lieb-Liniger-McGuire) \cite{LiebLiniger,McGuire}, realized in
\cite{Wolfgang_1D,Salomon_soliton,Hulet_soliton} in the mean-field
regime and in \cite{Tilman_1D,Bill_1D} in the regime of
interactions of intermediate strength; and
spin-$\frac{1}{2}$-fermions (Yang-Gaudin) \cite{Yang-Gaudin},
realized in \cite{Tilman_dimers}. The list has a potential to grow
to include also the fermionic $p$-wave version of hard-core
particles \cite{Doerte_p-waves,FTG}; the gas of $1/r^2$
interacting atoms (Calogero-Sutherland)
\cite{Calogero,Sutherland}; and the gas of fermions on a lattice
(Fermi-Hubbard) \cite{Fermi-Hubbard}. The experiment
\cite{Paredes_Nature}---which is a realization of the system whose
time-dynamics we study in the present paper---used an optical
lattice in the tight-binding regime
\cite{intro-B-H_opt_latt,expm-B-H_opt_latt}. The technique was
originally developed to reach the superfluid--Mott-insulator
transition \cite{intro_B-H} as achieved in \cite{expm-Mott}. We
should also mention the experimental studies
\cite{expm-1D_B-H_opt_latt} of a related nonintegrable system, the
one-dimensional lattice bosons with finite coupling.

An integrable model possesses many nontrivial integrals of motion,
and it is natural to wonder what consequences this fact may have
for time dynamics and kinetics. Perhaps the best known theoretical
efforts in this vein are the attempts to explain the suppression
of equilibration in the Fermi-Pasta-Ulam chains by the closeness
to various integrable models (see \cite{Izrailev_FPU_review} for a
review). Another research direction concerns the effects of integrals 
of motion on the autocorrelation properties of large systems, first 
studied in \cite{Mazur,Suzuki} and later specialized to spin systems
\cite{Prelovsek_conductance_1,Prelovsek_conductance_2}. More recent 
are the studies of the onset of thermalization in a large quantum system
\cite{Izrailev_thermalization_1,Izrailev_thermalization_2}, and in
particular in a mesoscopic-size Lieb-Liniger gas \cite{Izrailev_Bose}.

The major inspiration for our work---underlying especially Fig.
2 below---is the recent experiment on the non-equilibrium dynamics
of one-dimensional hard-core bosons in a harmonic potential
performed at Penn State University \cite{Weiss_thermalization}.
There it was found that hard-core bosons do not relax to the usual
state of thermodynamic equilibrium. The question that intrigued us
is whether nevertheless there exists \textit{some} kind of 
equilibrium state to which a many-body integrable system relaxes 
in the course of time evolution from even a highly nonequilibrium 
initial state---and, if so, how to predict mean values of physical
observables in such state.

{\it Generalized Gibbs ensemble}.-- We start with the latter
question, for now simply assuming that an equilibrium state exists. 
We conjecture that then the standard prescription of statistical
mechanics applies: one should maximize the many-body entropy $S =
k_{\mbox{\small B}}\,T\!r\left[\rho\ln(1/\rho)\right]$, subject to
the constraints imposed by all the integrals of motion. This
results in the following many-body density matrix:
\begin{eqnarray}
\hat{\rho}
 =Z^{-1}
  \exp\left[ -\sum_{m} \lambda_{m} \hat{{\cal I}}_{m} \right]
\quad,
\label{general_fully_constrained_rho}
\end{eqnarray}
where $\{\hat{{\cal I}}_{m}\}$ is the {\it full} set of the integrals of
motion, $Z=T\!r\left[\exp[ -\sum_{m} \lambda_{m}
\hat{{\cal I}}_{m}]\right]$ is the partition function, and
$\{\lambda_{m}\}$ are the Lagrange multipliers, fixed by the
initial conditions via
\begin{eqnarray}
T\!r\left[\hat{{\cal I}}_{m}\hat{\rho}\right] = \langle \hat{{\cal I}}_{m}\rangle(t=0)
\quad.
\label{general_constraints}
\end{eqnarray}
The generalized Gibbs ensemble
(\ref{general_fully_constrained_rho}) reduces to the usual
grand-canonical ensemble in the case of a generic system, where
the only integrals of motion are the total energy, the number of
particles, and, for periodic systems, the total momentum.
Conceptually, the ensemble (\ref{general_fully_constrained_rho})
is close to the one E.T.\ Jaynes introduced in 1957 in the context
of the so-called ``subjective statistical mechanics''
\cite{Jaynes}. Girardeau used Jaynes's concept to study the
relaxation of magnetization in the XY-model \cite{Marvin_XY}.
Below we test the predictive power of
(\ref{general_fully_constrained_rho}) and
(\ref{general_constraints}) on the example of hard-core bosons on
a one-dimensional lattice, a system integrable via Jordan-Wigner
mapping to free fermions.

{\it The Hamiltonian and the \mbox{(quasi-)}momentum distribution of hard-core
bosons on a lattice}.-- The Hamiltonian for hard core bosons (HCB)
on a one-dimensional lattice with $L$ sites reads
\begin{align}
&\quad\quad\quad\quad\quad\hat{H} = -J \sum_{i=1}^{L} \left(
\hat{b}^{\dagger}_{i} \hat{b}^{}_{i+1} + \mbox{h.c.} \right)
\label{HCB_Hamiltonian}
\\
\intertext{where}
&\begin{array}{lrlll} \!\!\!\!\!\!
[\hat{b}^{}_{i},\, \hat{b}^{\dagger}_{j}] = 0, &
   [\hat{b}^{}_{i},\, \hat{b}^{}_{j}]
&\!\!=&\!\!  [\hat{b}^{\dagger}_{i},\, \hat{b}^{\dagger}_{j}] =  0
& \mbox{for all $i$ and $j \ne i$;}
\\
\!\!\!\!\!\!\! \{ \hat{b}_{i},\, \hat{b}^\dagger_{i} \} = 1, &
   (\hat{b}^{}_{i})^2
&\!\!=&\!\!  (\hat{b}^{\dagger}_{i})^2 =  0 & \mbox{for all $i$.}
\end{array}
\nonumber
\end{align}
Here $\hat{b}^{}_{i}$ ($\hat{b}^\dagger_{i}$) is the annihilation
(creation) operator for hard-core bosons, and $J$ is the hopping
constant. 
For our theoretical predictions we use a
periodic lattice ($\hat{b}^{}_{L+1} = \hat{b}^{}_{1}$). However,
the subsequent numerical studies are performed for the more
experimentally relevant hard-wall boundary conditions. 
For sufficiently large lattice sizes $L$, the difference 
between real physical quantities calculated using these two 
settings is negligible \cite{lieb61}.

Our primary observable of interest is the HCB \mbox{(quasi-)}momentum
distribution $f(k) = \langle \hat{f}(k) \rangle$, normalized to
the total number of particles $N$, where
\begin{eqnarray}
\hat{f}(k) =
\frac{1}{L} \sum_{i=1}^{L} \sum_{i'=1}^{L}
e^{-i 2\pi k (i-i')/L} \hat{b}^{\dagger}_{i'}\hat{b}^{}_{i}
\label{HCB_momentum_distribution}
\end{eqnarray}
is the HCB \mbox{(quasi-)}momentum distribution operator.

{\it Fermi-Bose correspondence}.-- Our bosonic system can be mapped
to a free fermionic (FF) one via the Jordan-Wigner transformation
$
\hat{b}^{\dag}_i=
\hat{c}^{\dag}_i
\prod_{i'=1}^{i-1} e^{-i\pi \hat{c}^{\dag}_{i'} \hat{c}^{}_{i'}},\ \
\hat{b}_i=\prod_{i'=1}^{i-1} e^{i\pi
\hat{c}^{\dag}_{i'}\hat{c}^{}_{i'}} \hat{c}^{}_{i} $, where
$\hat{c}^{}_{i}$ ($\hat{c}^\dagger_{i}$) is the fermionic
annihilation (creation) operator. (Note that since the spatial
density operators for fermions and bosons are equal to each other,
$\hat{c}^{\dag}_{i}\hat{c}^{}_{i} =
\hat{b}^{\dag}_{i}\hat{b}^{}_{i}$, the inverse mapping is
straightforward.) Under this transformation our Hamiltonian
(\ref{HCB_Hamiltonian}) becomes just the Hamiltonian for
noninteracting fermions on a lattice:
\begin{align}
&\quad\quad\quad\quad\quad \hat{H} = -J \sum_{i=1}^{L} \left(
\hat{c}^{\dagger}_{i} \hat{c}^{}_{i+1} + \mbox{h.c.} \right),
\label{FF_Hamiltonian}
\\
&
\begin{array}{lcl}
\{\hat{c}^{}_{i},\, \hat{c}^{\dagger}_{j}\} = 1,\,
   \{\hat{c}^{}_{i},\, \hat{c}^{}_{j}\}
=  \{\hat{c}^{\dagger}_{i},\, \hat{c}^{\dagger}_{j}\} =  0 &
\mbox{for all $i$ and $j$}.
\end{array}
\nonumber
\end{align}
The corresponding fermions obey periodic (anti-periodic) boundary
conditions for odd (even) numbers of particles: $\hat{c}^{}_{L+1}
= (-1)^{\hat{N}+1}\hat{c}^{}_{1}$. Here and below $\hat{N} =
\sum_{i=1}^{L} \hat{c}^\dagger_{i} \hat{c}^{}_{i} = \sum_{i=1}^{L}
\hat{b}^\dagger_{i} \hat{b}^{}_{i}$ is the particle number
operator, the same for both fermions and bosons.

{\it Integrals of motion}.-- It is clear from the fermionic form
(\ref{FF_Hamiltonian}) of the Hamiltonian that our system
possesses as many conserved quantities as there are lattice sites:
they are simply the fermionic \mbox{(quasi-)}momentum distribution operators
\begin{eqnarray}
\hat{{\cal I}}_{k}
= \hat{f}^{F}(k)
=
\frac{1}{L} \sum_{i=1}^{L} \sum_{i'=1}^{L}
\sigma_{i-i'}(\hat{N})
e^{-i 2\pi k (i-i')/L}
\hat{c}^{\dagger}_{i'}\hat{c}^{}_{i}
\label{I-integrals of motion}
\end{eqnarray}
where $\sigma_{\Delta_i}(N) = 1$ for odd $N$, and 
$\sigma_{\Delta_i}(N)= e^{-i \pi \Delta_i/L}$ for even $N$.

Note that if expressed through the bosonic fields, the above
integrals of motion become complicated many-body operators.
Consider, for example, the lattice analog of the fourth moment of
the fermionic \mbox{(quasi-)}momentum distribution $\hat{I}_{4}$, defined as
\begin{eqnarray}
&&\frac{1}{4}\left(\frac{2\pi}{L}\right)^{4} \hat{I}_{4} = \sum_{k}
\left(1-\cos \frac{2\pi n}{L}\right)^{2} \hat{f}^{F}(k) \\&&=
   \frac{3}{2}\hat{N} + \frac{1}{J}\hat{H} +
    \frac{1}{4} \sum_{i=1}^{L}
               \left(
    \hat{b}^\dagger_{i}
    (
    1-2 \hat{b}^\dagger_{i+1} \hat{b}^{}_{i+1}
    )
    \,\hat{b}^{}_{i+2}
    + \mbox{h.c.}
              \right). \nonumber
\end{eqnarray} 
It is one of the simplest linear combinations of the integrals
of motion (\ref{I-integrals of motion}), but it becomes a two-body operator in the bosonic
representation.

\begin{figure}
%
\includegraphics[scale=.55]{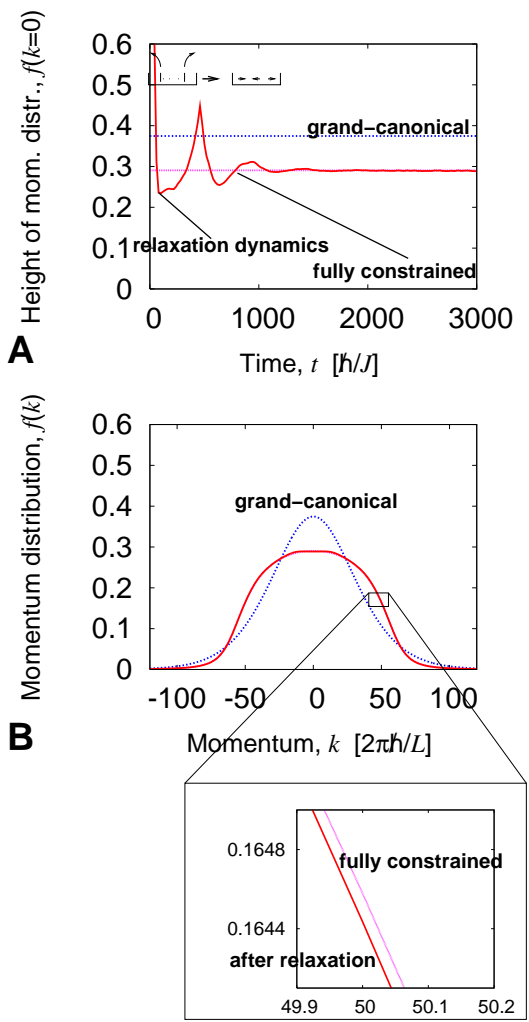}
\caption { \label{f:momentum} (color online). Momentum distribution of $N=30$
hard-core bosons undergoing a free expansion from an initial
zero-temperature hard-wall box of size $L_{\mbox{\small in.}} =
150$ to the final hard-wall box of size $L=600$. The initial box
is situated in the middle of the final one. (a) Approach to
equilibrium. (b) Equilibrium \mbox{(quasi-)}momentum distribution after
relaxation in comparison with the predictions of the
grand-canonical and of the fully constrained
(\ref{fully_constrained_rho}) thermodynamical ensembles. The
prediction of the fully constrained ensemble is virtually
indistinct from the results of the dynamical simulation; see the
inset for a measure of the accuracy. (An animation of the time
evolution is posted on line \cite{movies}.)
         }
\end{figure}

{\it Fully constrained thermodynamic ensemble}.-- The density matrix for the
fully constrained thermodynamic ensemble described above reads
\begin{eqnarray}
\hat{\rho}_{\mbox{\small f.c.}}
 =Z_{\mbox{\small f.c.}}^{-1}
  \exp\left[ -\sum_{k} \lambda_{k} \hat{f}^{F}(k) \right]
\quad,
\label{fully_constrained_rho}
\end{eqnarray}
where $Z_{\mbox{\small f.c.}}= T\!r\left[\exp[ -\sum_{k}
\lambda_{k} \hat{f}^{F}(k)]\right] = \prod_{k}
(1+e^{-\lambda_{k}})$. The values of the Lagrange multipliers
$\lambda_{k}$ must be fixed by the requirement that the fermionic
\mbox{(quasi-)}momentum distribution predicted by (\ref{fully_constrained_rho})
be the same as the \mbox{(quasi-)}momentum distribution of fermions in the actual
initial---or, for that matter, time-evolved---state of the system.
This constraint leads to $ \lambda_{k} =
\ln\left(\frac{1-f^{F}(k)}{f^{F}(k)}\right)$. As we stated above,
the density matrix given by (\ref{fully_constrained_rho}) is
assumed to predict correctly the values of the system's
observables after a complete relaxation from an initial state with
the fermionic \mbox{(quasi-)}momentum distribution given by $f^{F}(k)$. Below, we
test this conjecture numerically, using the bosonic \mbox{(quasi-)}momentum
distribution as the figure of merit.

\begin{figure}
%
\includegraphics[scale=.54]{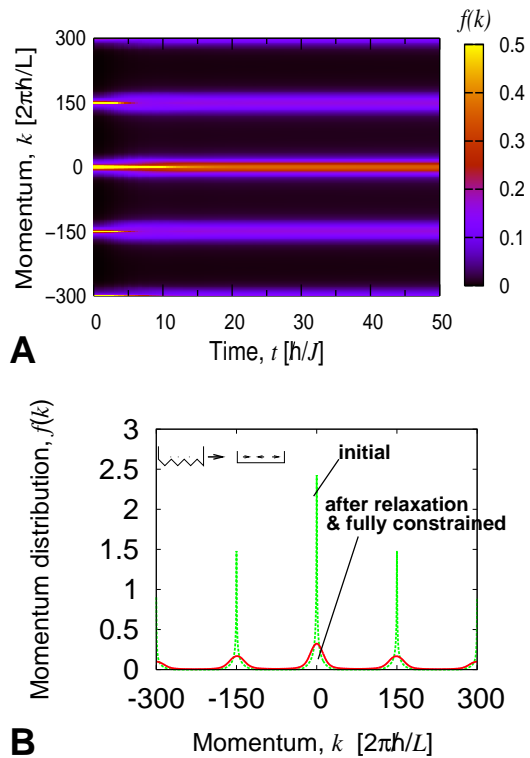}
\caption { \label{f:superlattice} (color online). Time evolution of 
the \mbox{(quasi-)}momentum distribution (a) and the 
\mbox{(quasi-)}momentum distribution after relaxation
(b) of $N=30$ hard-core bosons undergoing a free expansion from an
initial zero-temperature superlattice with period four of
half-depth $A=8J$ and bound by a hard-wall box of size $L=600$,
to the final flat-bottom box ($A=0$) of the same size. The discrepancy
between the result of time propagation and the prediction of the
fully constrained ensemble (\ref{fully_constrained_rho}) (also
shown in (b)) is less than the width of the line. Momentum peaks
remain well-resolved during the whole duration of propagation;
$t_{fin.} = 3000 \hbar/J$ for the subfigure (b). (An animation for
the time evolution can also be found in \cite{movies}.)
         }
\end{figure}

{\it Numerical tests}.-- In order to verify our predictions we
perform two series of text-book-like numerical experiments on the
relaxation of an ensemble of hard-core bosons on a lattice from a
highly nonequilibrium state. We have chosen to study lattices in 
which the final size $L>>N$, i.e., the average interparticle 
distance is much larger than the lattice spacing, so that our results 
are also of relevance to continuous systems \cite{rigol04_1}. 
The numerical technique has been described elsewhere \cite{time_prop-lattice}. 

In the first series we prepare our gas in the ground state of a
hard-wall box, then let the gas expand to a larger box. For all
sizes of the initial box, we find that the \mbox{(quasi-)}momentum 
distribution indeed converges to an almost time independent
distribution (see Figure \ref{f:momentum}). Next, we compare 
the result after relaxation with the predictions of standard 
statistical mechanics and of the fully constrained ensemble
(\ref{general_fully_constrained_rho}), (\ref{fully_constrained_rho}).
We find that the fully constrained thermodynamics stands in an
exceptional agreement with the results of the dynamical
propagation. (See \cite{therm_Rigol} for further details of the
thermal algorithm.) The accuracy of the above predictions has been
successfully verified for the whole range of available values of
the size of the initial box, from $L_{\mbox{\small in.}} = N = 30$
through $L_{\mbox{\small in.}} = L = 600$.

In the second series (Figure \ref{f:superlattice}) we study the
effects of the memory of the initial conditions that is stored in
the fully constrained ensemble
(\ref{general_fully_constrained_rho}), (\ref{fully_constrained_rho}).
Our setting is very similar to an actual experiment on relaxation
of an ensemble of hard-core bosons in a harmonic potential
\cite{Weiss_thermalization}. There the momentum distribution was
initially split into two peaks. After many periods of oscillation,
no appreciable relaxation to a single-bell distribution was
observed. In our case the system is initially in the ground state
of a hard-wall box with a superlattice (spatially-periodic background 
potential, see \cite{superlattice} for details) with period 4, 
\begin{equation}
\label{addpot}
  \hat{V}_{\rm ext} = A \sum_{i} {\rm cos} \frac{2 \pi i}{T} \,\, 
  \hat{b}^{\dagger}_{i} \hat{b}^{}_{i},
  \quad T=4
\end{equation}
and is subsequently released to a flat-bottom hard-wall box $\hat{V}_{\rm ext}=0$. 
Our results show that even after a very long propagation time, the four
characteristic peaks in the \mbox{(quasi-)}momentum distribution remain well
resolved, although their shape is modified in the course of the
propagation. Our interpretation of both experimental and numerical
results is as follows: if the initial \mbox{(quasi-)}momentum distribution
consists of several well-separated peaks, the memory of the
initial distribution that is stored in the ensemble
(\ref{general_fully_constrained_rho}) prevents the peaks from
overlapping, no matter how long the propagation time. Note also
that the residual broadening of the peaks seen in Figure
\ref{f:superlattice} is beyond the experimental accuracy in
\cite{Weiss_thermalization}.

{\it Summary.}-- We have demonstrated that an integrable many-body
quantum system---one-dimensional hard-core bosons on a
lattice---can undergo relaxation to an equilibrium
state. The properties of this state are governed by the usual laws
of statistical physics, properly updated to accommodate all the
integrals of motion. We further show that our generalized
equilibrium state carries more memory of the initial conditions
than the usual thermodynamic one. It is in the light of that
observation that we interpret the results of the recent experiment
on the non-equilibrium dynamics of one-dimensional hard-core
bosons performed at Penn State University
\cite{Weiss_thermalization}, where an initial two-peaked \mbox{(quasi-)}momentum
distribution failed to relax to a single-bell distribution.

\begin{acknowledgments}
We are grateful to David Weiss, Boris Svistunov, Hubert Saleur, 
Tommaso Roscilde, Rajiv R. P. Singh, and Marvin Girardeau
for enlightening discussions on the subject. This work was 
supported by a grant from Office of Naval Research No.
N00014-03-1-0427, by grants from the National Science 
Foundation No. PHY-0301052, No. DMR-0240918, No. DMR-0312261,
and through the National Science Foundation grant for the 
Institute for Theoretical Atomic and Molecular Physics at 
Harvard University and Smithsonian Astrophysical Observatory.
\end{acknowledgments}

%

\end{document}